\documentclass[12pt,a4paper]{article}

\usepackage{rotating}  %
\usepackage{graphpap}  %
\usepackage{multirow}  %
\usepackage{subfigure} %
\usepackage{cancel}    %

\usepackage{longtable} 

\usepackage{ifthen} 
\newboolean{pdflatex}
\setboolean{pdflatex}{true} 

\newboolean{articletitles}
\setboolean{articletitles}{true} 

\newboolean{uprightparticles}
\setboolean{uprightparticles}{true} 

\usepackage{microtype}
\usepackage{lineno}  
\usepackage{xspace} 

\usepackage{graphicx}  
\usepackage{color}
\usepackage{colortbl}
\graphicspath{{./figs/}} 

\usepackage{amsmath} 
\usepackage{amssymb}
\usepackage{amsfonts}
\usepackage{upgreek} 

\newcommand*\patchAmsMathEnvironmentForLineno[1]{%
\expandafter\let\csname old#1\expandafter\endcsname\csname #1\endcsname
\expandafter\let\csname oldend#1\expandafter\endcsname\csname
end#1\endcsname
 \renewenvironment{#1}%
   {\linenomath\csname old#1\endcsname}%
   {\csname oldend#1\endcsname\endlinenomath}%
}
\newcommand*\patchBothAmsMathEnvironmentsForLineno[1]{%
  \patchAmsMathEnvironmentForLineno{#1}%
  \patchAmsMathEnvironmentForLineno{#1*}%
}
\AtBeginDocument{%
\patchBothAmsMathEnvironmentsForLineno{equation}%
\patchBothAmsMathEnvironmentsForLineno{align}%
\patchBothAmsMathEnvironmentsForLineno{flalign}%
\patchBothAmsMathEnvironmentsForLineno{alignat}%
\patchBothAmsMathEnvironmentsForLineno{gather}%
\patchBothAmsMathEnvironmentsForLineno{multline}%
}

\usepackage{hyperref}    
\usepackage[all]{hypcap} 

\def\lhcb {\mbox{LHCb}\xspace}
\def\ux85 {\mbox{UX85}\xspace}

\ifthenelse{\boolean{uprightparticles}}%
{
 
 \def\Pgamma      {\ensuremath{\upgamma}\xspace}

 \def\Pmu         {\ensuremath{\upmu}\xspace}                 
 \def\Pnu         {\ensuremath{\upnu}\xspace}                 
                  
 \def\Ppi         {\ensuremath{\uppi}\xspace}

 \def\Pphi        {\ensuremath{\upphi}\xspace}

 \def\Ppsi        {\ensuremath{\uppsi}\xspace}

 \def\PDelta      {\ensuremath{\Delta}\xspace}                 
 \def\PXi      {\ensuremath{\Xi}\xspace}                 
 \def\PLambda      {\ensuremath{\Lambda}\xspace}                 
 \def\PSigma      {\ensuremath{\Sigma}\xspace}                 
 \def\POmega      {\ensuremath{\Omega}\xspace}                 
 \def\PUpsilon      {\ensuremath{\Upsilon}\xspace}                 
 
 \def\PB      {\ensuremath{\mathrm{B}}\xspace}                 
                  
 \def\PD      {\ensuremath{\mathrm{D}}\xspace}

 \def\PJ      {\ensuremath{\mathrm{J}}\xspace}                 
 \def\PK      {\ensuremath{\mathrm{K}}\xspace}

 \def\Pb      {\ensuremath{\mathrm{b}}\xspace}                 
 \def\Pc      {\ensuremath{\mathrm{c}}\xspace}

 \def\Pi      {\ensuremath{\mathrm{i}}\xspace}

 \def\Ps      {\ensuremath{\mathrm{s}}\xspace}

}
{
 
 \def\Pgamma      {\ensuremath{\gamma}\xspace}

 \def\Pmu         {\ensuremath{\mu}\xspace}                 
 \def\Pnu         {\ensuremath{\nu}\xspace}                 
                  
 \def\Ppi         {\ensuremath{\pi}\xspace}

 \def\Pphi        {\ensuremath{\phi}\xspace}

 \def\Ppsi        {\ensuremath{\psi}\xspace}                 
                  
 \mathchardef\PDelta="7101
 \mathchardef\PXi="7104
 \mathchardef\PLambda="7103
 \mathchardef\PSigma="7106
 \mathchardef\POmega="710A
 \mathchardef\PUpsilon="7107
                  
 \def\PB      {\ensuremath{B}\xspace}                 
                  
 \def\PD      {\ensuremath{D}\xspace}

 \def\PJ      {\ensuremath{J}\xspace}                 
 \def\PK      {\ensuremath{K}\xspace}

 \def\Pb      {\ensuremath{b}\xspace}                 
 \def\Pc      {\ensuremath{c}\xspace}

 \def\Pi      {\ensuremath{i}\xspace}

 \def\Ps      {\ensuremath{s}\xspace}

}

\def\mumu       {\ensuremath{\Pmu^+\Pmu^-}\xspace}

\def\squark    {\ensuremath{\Ps}\xspace}

\def\cquark    {\ensuremath{\Pc}\xspace}

\def\bquark    {\ensuremath{\Pb}\xspace}

\def\pion  {\ensuremath{\Ppi}\xspace}

\def\pip   {\ensuremath{\pion^+}\xspace}
\def\pim   {\ensuremath{\pion^-}\xspace}
\def\pipi  {\ensuremath{\pion^+\pion^-}\xspace}

\def\kaon  {\ensuremath{\PK}\xspace}
  \def\Kbar  {\kern 0.2em\overline{\kern -0.2em \PK}{}\xspace}

\def\Kz    {\ensuremath{\kaon^0}\xspace}
\def\Kzb   {\ensuremath{\Kbar^0}\xspace}
\def\KzKzb {\ensuremath{\Kz \kern -0.16em \Kzb}\xspace}
\def\Kp    {\ensuremath{\kaon^+}\xspace}
\def\Km    {\ensuremath{\kaon^-}\xspace}

\def\KpKm  {\ensuremath{\Kp \kern -0.16em \Km}\xspace}

\def\Dbar    {\kern 0.1em\overline{\kern -0.1em \PD}{}\xspace}
\def\D       {\ensuremath{\PD}\xspace}

\def\Dz      {\ensuremath{\D^0}\xspace}
\def\Dzb     {\ensuremath{\Dbar^0}\xspace}
\def\DzDzb   {\ensuremath{\Dz {\kern -0.16em \Dzb}}\xspace}
\def\Dp      {\ensuremath{\D^+}\xspace}
\def\Dm      {\ensuremath{\D^-}\xspace}

\def\DpDm    {\ensuremath{\Dp {\kern -0.16em \Dm}}\xspace}

\def\Ds      {\ensuremath{\D^+_\squark}\xspace}

\def\B       {\ensuremath{\PB}\xspace}
\def\Bbar    {\ensuremath{\kern 0.18em\overline{\kern -0.18em \PB}{}}\xspace}

\def\Bz      {\ensuremath{\B^0}\xspace}

\def\Bu      {\ensuremath{\B^+}\xspace}

\def\Bp      {\ensuremath{\Bu}\xspace}

\def\Bd      {\ensuremath{\B^0}\xspace}

\def\Bc      {\ensuremath{\B_\cquark^+}\xspace}

\def\jpsi     {\ensuremath{{\PJ\mskip -3mu/\mskip -2mu\Ppsi\mskip 2mu}}\xspace}
\def\psitwos  {\ensuremath{\Ppsi{(2\mathrm{S})}}\xspace}

  \def\Y#1S{\ensuremath{\PUpsilon{(#1S)}}\xspace}

\def\Lbar {\ensuremath{\kern 0.1em\overline{\kern -0.1em\PLambda}}\xspace}

\def\BF         {{\ensuremath{\cal B}\xspace}}

\def\BR         {\BF}

\def\to                 {\ensuremath{\rightarrow}\xspace}

\def\order   {\ensuremath{\mathcal{O}}\xspace}

\def\AT#1     {\ensuremath{A_{\mathrm{T}}^{#1}}\xspace}           

\def\C#1      {\ensuremath{\mathcal{C}_{#1}}\xspace}                       
\def\Cp#1     {\ensuremath{\mathcal{C}_{#1}^{'}}\xspace}                    
\def\Ceff#1   {\ensuremath{\mathcal{C}_{#1}^{\mathrm{(eff)}}}\xspace}        
\def\Cpeff#1  {\ensuremath{\mathcal{C}_{#1}^{'\mathrm{(eff)}}}\xspace}       
\def\Ope#1    {\ensuremath{\mathcal{O}_{#1}}\xspace}                       
\def\Opep#1   {\ensuremath{\mathcal{O}_{#1}^{'}}\xspace}                    

\newcommand{\tev}{\ensuremath{\mathrm{\,Te\kern -0.1em V}}\xspace}
\newcommand{\gev}{\ensuremath{\mathrm{\,Ge\kern -0.1em V}}\xspace}
\newcommand{\mev}{\ensuremath{\mathrm{\,Me\kern -0.1em V}}\xspace}
\newcommand{\kev}{\ensuremath{\mathrm{\,ke\kern -0.1em V}}\xspace}
\newcommand{\ev}{\ensuremath{\mathrm{\,e\kern -0.1em V}}\xspace}
\newcommand{\gevc}{\ensuremath{{\mathrm{\,Ge\kern -0.1em V\!/}c}}\xspace}
\newcommand{\mevc}{\ensuremath{{\mathrm{\,Me\kern -0.1em V\!/}c}}\xspace}
\newcommand{\gevcc}{\ensuremath{{\mathrm{\,Ge\kern -0.1em V\!/}c^2}}\xspace}
\newcommand{\gevgevcccc}{\ensuremath{{\mathrm{\,Ge\kern -0.1em V^2\!/}c^4}}\xspace}
\newcommand{\mevcc}{\ensuremath{{\mathrm{\,Me\kern -0.1em V\!/}c^2}}\xspace}

\def\mm   {\ensuremath{\rm \,mm}\xspace}

\def\mum  {\ensuremath{\,\upmu\rm m}\xspace}

\def\invfb   {\ensuremath{\mbox{\,fb}^{-1}}\xspace}

\newcommand{\stat}{\ensuremath{\mathrm{(stat)}}\xspace}
\newcommand{\syst}{\ensuremath{\mathrm{(syst)}}\xspace}

\def\order{{\ensuremath{\cal O}}\xspace}

\def\gsim{{~\raise.15em\hbox{$>$}\kern-.85em
          \lower.35em\hbox{$\sim$}~}\xspace}
\def\lsim{{~\raise.15em\hbox{$<$}\kern-.85em
          \lower.35em\hbox{$\sim$}~}\xspace}

\def\sPlot{\mbox{\em sPlot}}

\def\sqs   {\ensuremath{\protect\sqrt{s}}\xspace}

\def\pt         {\mbox{$p_{\rm T}$}\xspace}

\def\evtgen     {\mbox{\textsc{EvtGen}}\xspace}
\def\pythia     {\mbox{\textsc{Pythia}}\xspace}

\def\geant      {\mbox{\textsc{Geant4}}\xspace}

\def\photos     {\mbox{\textsc{Photos}}\xspace}

\def\tell1  {TELL1\xspace}
\def\ukl1   {UKL1\xspace}

\usepackage{cite} 
\usepackage{mciteplus}

\newcommand{\Dsst}     {\ensuremath{\D{}^{\ast+}_{\mathrm{s}}}\xspace}
\newcommand{\BctoDs}   {\ensuremath{\Bc{}\to{}\jpsi{}\Ds}\xspace}
\newcommand{\BctoDsst} {\ensuremath{\Bc{}\to{}\jpsi{}\Dsst}\xspace}

\newcommand{\BctoPPP}  {\ensuremath{\Bc{}\to{}\jpsi{}\pip{}\pip{}\pim}\xspace}
\newcommand{\BctoP}    {\ensuremath{\Bc{}\to{}\jpsi{}\pip}\xspace}

\newcommand{\jpsimm}   {\ensuremath{\jpsi{}\to{}\mumu}\xspace}
\newcommand{\dsphipi}  {\ensuremath{\Ds{}\to\left(\Kp{}\Km{}\right)_{\Pphi}\pip}\xspace}

\newcommand{\bctods}   {\BctoDs}
\newcommand{\bctodsst} {\BctoDsst}
\newcommand{\bctoppp}  {\BctoPPP}
\newcommand{\bctop}    {\BctoP}
\newcommand{\bctopi}   {\BctoP}

\newcommand{\rdspi} {\ensuremath{\mathcal{R}_{\Ds\mskip -6mu/\pip}}}
\newcommand{\rdsds} {\ensuremath{\mathcal{R}_{\Dsst\mskip -6mu/\Ds}}}

\renewcommand{\Dbar}{\ensuremath{\overline{\D}}\xspace}

\begin{document}

\renewcommand{\thefootnote}{\fnsymbol{footnote}}
\setcounter{footnote}{1}




\begin{titlepage}
\pagenumbering{roman}

\vspace*{-1.5cm}
\centerline{\large EUROPEAN ORGANIZATION FOR NUCLEAR RESEARCH (CERN)}
\vspace*{1.5cm}
\hspace*{-0.5cm}
\begin{tabular*}{\linewidth}{lc@{\extracolsep{\fill}}r}
\\
 & & CERN-PH-EP-2013-051 \\  
 & & LHCb-PAPER-2013-010 \\  
 & & \\
\end{tabular*}

\vspace*{1.5cm}

{\bf\boldmath\huge
\begin{center}
  Observation of 
      {\boldmath$\Bc{}\to{}\jpsi{}\D{}^{+}_{\mathrm{s}}$} 
  and {\boldmath$\Bc{}\to{}\jpsi{}\D{}^{\ast+}_{\mathrm{s}}$} 
  decays
\end{center}
}

\vspace*{1.0cm}

\begin{center}
The LHCb collaboration\footnote{Authors are listed on the following pages.}
\end{center}

\vspace{\fill}

\begin{abstract}
  \noindent
  The decays \bctods~and \BctoDsst~are observed for the first time  
  using a dataset, corresponding to an integrated luminosity
  of 3\invfb, collected by the LHCb experiment in 
  proton-proton collisions at centre-of-mass energies of \sqs= 7~and~8\tev. 
  The statistical significance for both signals is in excess of
  9~standard deviations.
  The following ratios of branching fractions are measured to be
  \begin{eqnarray*}
    \dfrac {\BR \left( \bctods \right) } 
           {\BR \left( \bctop  \right) } 
           & = & 2.90 \pm 0.57 \pm 0.24,  \\ 
   \dfrac {  \BR\left( \BctoDsst \right) }
          {  \BR\left( \BctoDs   \right) } 
          & = &  2.37 \pm 0.56 \pm 0.10,
  \end{eqnarray*}
  where the first uncertainties are statistical and the second systematic.

The mass of the~\Bc~meson is measured to be 
\begin{equation*}
  m_{\Bc} = 6276.28 \pm 1.44\,\stat \pm 0.36\,\syst\mevcc,
\end{equation*}
using the \bctods~decay mode. 
\end{abstract}

\vspace*{1.0cm}

\begin{center}
  Published in Physical Review D. 
\end{center}

\vspace{\fill}

{\footnotesize 
\centerline{\copyright~CERN on behalf of the \lhcb collaboration, license \href{http://creativecommons.org/licenses/by/3.0/}{CC-BY-3.0}.}}
\vspace*{2mm}

\end{titlepage}


\newpage
\setcounter{page}{2}
\mbox{~}
\newpage

\centerline{\large\bf LHCb collaboration}
\begin{flushleft}
\small
R.~Aaij$^{40}$, 
C.~Abellan~Beteta$^{35,n}$, 
B.~Adeva$^{36}$, 
M.~Adinolfi$^{45}$, 
C.~Adrover$^{6}$, 
A.~Affolder$^{51}$, 
Z.~Ajaltouni$^{5}$, 
J.~Albrecht$^{9}$, 
F.~Alessio$^{37}$, 
M.~Alexander$^{50}$
% ...

{\footnotesize \it
% ...
$ ^{5}$Clermont Universit\'{e}, Universit\'{e} Blaise Pascal, CNRS/IN2P3, LPC, Clermont-Ferrand, France\\
$ ^{6}$CPPM, Aix-Marseille Universit\'{e}, CNRS/IN2P3, Marseille, France\\
% ...
$ ^{9}$Fakult\"{a}t Physik, Technische Universit\"{a}t Dortmund, Dortmund, Germany\\
% ...
$ ^{35}$Universitat de Barcelona, Barcelona, Spain\\
$ ^{36}$Universidad de Santiago de Compostela, Santiago de Compostela, Spain\\
$ ^{37}$European Organization for Nuclear Research (CERN), Geneva, Switzerland\\
% ...
$ ^{40}$Nikhef National Institute for Subatomic Physics, Amsterdam, The Netherlands\\
% ...
$ ^{45}$H.H. Wills Physics Laboratory, University of Bristol, Bristol, United Kingdom\\
% ...
$ ^{50}$School of Physics and Astronomy, University of Glasgow, Glasgow, United Kingdom\\
$ ^{51}$Oliver Lodge Laboratory, University of Liverpool, Liverpool, United Kingdom\\
% ...
$ ^{n}$LIFAELS, La Salle, Universitat Ramon Llull, Barcelona, Spain\\
% ...
}
\end{flushleft}

\cleardoublepage


\renewcommand{\thefootnote}{\arabic{footnote}}
\setcounter{footnote}{0}



\pagestyle{plain} 
\setcounter{page}{1}
\pagenumbering{arabic}


%

\clearpage


\section{Introduction}
\label{seq:intro}
The \Bc~meson, the ground state of  the $\bar{\bquark}{}\cquark$~system, 
is unique, being the only weakly decaying heavy quarkonium
system. Its lifetime~\cite{PDG2012,Aaltonen:2012yb} is 
almost three times smaller than that of other beauty 
mesons, pointing to the important role of the charm quark in weak 
\Bc~decays.  
The~\Bc meson was first observed through its semileptonic decay 
\mbox{$\Bc\to\jpsi\ell^+\Pnu_{\ell}\mathrm{X}$}~\cite{Abe:1998wi}.  
Only three hadronic modes have 
been observed so far: 
\mbox{\bctop}~\cite{Abulencia:2005usa}, 
\mbox{\bctoppp}~\cite{LHCb-PAPER-2011-044}
and \mbox{$\Bc{}\to{}\psitwos{}\pip$}~\cite{LHCb-PAPER-2012-054}. 

\begin{figure}[htb]
  \setlength{\unitlength}{1mm}
  \centering
  \includegraphics*[width=150mm,height=40mm%
  ]{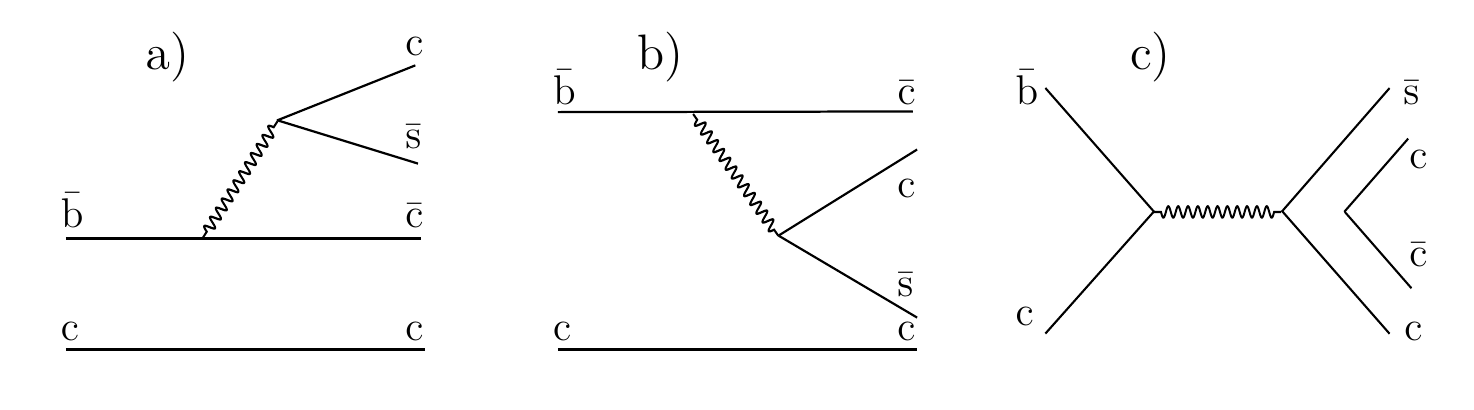}
  \caption { \small
    Feynman diagrams for \BctoDs decays: 
    (a)~spectator, 
    (b)~colour-suppressed spectator and (c)~annihilation topology.}
  \label{fig:diagrams}
\end{figure}

The first observations of the decays \bctods~and 
\BctoDsst~are reported in this paper. 
The leading Feynman diagrams of these decays are shown 
in Fig.~\ref{fig:diagrams}.
The decay \BctoDs is expected to proceed mainly through spectator
and colour-suppressed spectator diagrams. In contrast to
decays of other beauty hadrons, the weak annihilation topology is 
not suppressed and can contribute significantly to the decay amplitude.

Assuming that the spectator diagram dominates 
and that factorization holds, 
the following approximations can be established 
\begin{subequations}
\begin{eqnarray}
\rdspi
& \equiv &     
\dfrac {\Gamma \left( \bctods \right) }
       {\Gamma \left( \bctopi \right) } 
\approx 
\dfrac {\Gamma \left( \B{}\to{}\Dbar{}^{\ast}{}\Ds  \right) }
       {\Gamma \left( \B{}\to{}\Dbar{}^{\ast}{}\pip \right) }, \label{eq:exp_fact}  \\ 
\rdsds
& \equiv &    
\dfrac {\Gamma \left( \BctoDsst \right) }
       {\Gamma \left( \bctods   \right) } 
\approx 
\dfrac {\Gamma \left( \B{}\to{}\Dbar{}^{\ast}{}\D{}^{\ast+}_{\mathrm{s}}  \right) }
       {\Gamma \left( \B{}\to{}\Dbar{}^{\ast}{}\Ds \right) }, \label{eq:exp_fact2}  
\end{eqnarray}\label{eq:fact}
\end{subequations}
where \B~stands for \Bp~or $\B^{0}$ and 
$\Dbar{}^{\ast}$~denotes $\Dbar^{\ast0}$~or $\D^{\ast-}$.
Phase space corrections amount to $\order(0.5\%)$ for Eq.~\eqref{eq:exp_fact} 
and can be as large as 28\% for Eq.~\eqref{eq:exp_fact2},
depending on the relative orbital momentum.   
The relative branching ratios estimated in this way,
together with more detailed theoretical calculations, 
are listed in Table~\ref{tab:theory},
where the branching fractions for the $\B{}\to{}\Dbar{}^{\ast}{}\Ds$~and 
$\B{}\to{}\Dbar{}^{\ast}{}\pip$~decays are taken from Ref.~\cite{PDG2012}.

\begin{table}[htb]
  \centering
  \caption{ \small 
    Predictions for the ratios of \Bc~meson branching fractions. In
    the case of \rdsds\   the~second uncertainty is related 
    to the unknown relative orbital momentum.
  } \label{tab:theory}
  \vspace*{3mm}
  \begin{tabular*}{0.9\textwidth}{@{\hspace{10mm}}c@{\extracolsep{\fill}}cc@{\hspace{10mm}}}
      \rdspi
    & \rdsds
    & 
   \\
   \hline
    $2.90 \pm 0.42$
   & $2.20 \pm 0.35 \pm 0.62$
   & Eqs.~\eqref{eq:fact} with \Bd
   \\
     $1.58 \pm 0.34$
   & $2.07 \pm 0.52 \pm 0.52$
   & Eqs.~\eqref{eq:fact} with \Bu
   \\
     1.3 
   & 3.9 
   & Ref.~\,\cite{Kiselev:2003mp}
   \\
     2.6 
   & 1.7 
   & Ref.~\,\cite{Colangelo:1999zn}
   \\ 
     2.0    
   & 2.9    
   & Ref.~\,\cite{Ivanov:2006ni} 
   \\
     2.2  
   &  ---
   & Ref.~\cite{Dhir:2008hh} 
   \\
     1.2    
   &  ---  
   & Ref.~\cite{Chang:1992pt} 
    \\
 \end{tabular*}   
\end{table}

The analysis presented here 
is based on a data sample, corresponding to an integrated luminosity of 3\invfb,
collected with the LHCb detector during 2011 and 2012 in $\mathrm{pp}$~collisions 
at centre-of-mass  energies of 7~and~8\tev, respectively. 
The decay \mbox{\bctop} is used as a~normalization channel 
for the measurement of the branching fraction $\mbox{\BR(\bctods)}$. 
In~addition, the low energy release ($Q$-value) in the
$\bctods$~mode allows a determination of the \Bc mass 
with small systematic uncertainty.

\section{\lhcb detector}
\label{sec:Detector} 
The \lhcb detector~\cite{Alves:2008zz} is a single-arm forward
spectrometer covering the \mbox{pseudorapidity} range $2<\eta <5$,
designed for the study of particles containing \bquark or \cquark
quarks. The detector includes a high precision tracking system
consisting of a silicon-strip vertex detector surrounding the 
$\mathrm{pp}$~interaction region, a large-area silicon-strip detector located
upstream of a dipole magnet with a bending power of about
$4{\rm\,Tm}$, and three stations of silicon-strip detectors and straw
drift tubes placed downstream. The combined tracking system has 
momentum resolution $\Delta p/p$ that varies from 0.4\% at 5\gevc to
0.6\% at 100\gevc, and impact parameter resolution of 20\mum for
tracks with high transverse momentum. Charged hadrons are identified
using two ring-imaging Cherenkov detectors. Photon, electron and
hadron candidates are identified by a calorimeter system consisting of
scintillating-pad and preshower detectors, an electromagnetic
calorimeter and a hadronic calorimeter. Muons are identified by a
system composed of alternating layers of iron and multiwire
proportional chambers. The trigger~\cite{Aaij:2012me} consists of a
hardware stage, based on information from the~calorimeter and muon
systems, followed by a software stage which applies a full event
reconstruction. 

This analysis uses events collected by triggers that select the decay
products of the~dimuon decay of the $\jpsi$ meson with high
efficiency. At the hardware stage either one or two identified muon
candidates are required. In the case of single
muon triggers the transverse momentum, \pt, of the candidate is
required to be larger than 1.5\gevc. For dimuon candidates 
a requirement on the product of the \pt~of the muon candidates
is applied,  \mbox{$\sqrt{\pt_1 \pt_2} > 1.3\gevc$}. 
At the subsequent software trigger stage, two muons with 
invariant mass in the interval \mbox{$2.97 < m_{\mumu} <3.21\gevcc$} 
and consistent with originating from a common vertex
are required.

The detector acceptance and response are estimated with simulated data. 
Proton-proton~collisions are generated using
\pythia~6.4~\cite{Sjostrand:2006za} with the configuration described
in Ref.~\cite{LHCb-PROC-2010-056}. 
Particle decays are then simulated by \evtgen~\cite{Lange:2001uf} 
in which final state
radiation is generated using \photos~\cite{Golonka:2005pn}. The
interaction of the generated particles with the detector and its
response are implemented using the \geant
toolkit~\cite{Allison:2006ve,*Agostinelli:2002hh} as described in
Ref.~\cite{LHCb-PROC-2011-006}.

%
%
\section{Event selection} 
\label{seq:evsel}
Track quality of charged particles is ensured by requiring that 
the $\chi^2$ per degree
of freedom, $\chi^2_{\rm{tr}}/\mathrm{ndf}$, is less than~$4$. 
Further suppression of fake tracks created by the reconstruction 
is achieved by a neural network trained to 
discriminate between these and real particles
based on information from track fit and hit 
pattern in the tracking detectors. 
A~requirement on the output of this neural network, 
$\mathcal{P}_{\mathrm{fake}}<0.5$ allows to reject half of the fake tracks.

Duplicate particles created
by the reconstruction are suppressed by requiring the 
symmetrized  Kullback-Leibler 
divergence~\cite{LHCb-2008-002,*Kullback1,*Kullback3}, $\Delta^{\rm min}_{\rm KL}$, 
calculated with respect to all particles in the event, to be in excess
of 5000. In addition, the transverse momentum is required to be
greater than 550 (250)\mevc~for each muon~(hadron) candidate. 

Well identified muons are selected by requiring that
the difference in logarithms of the~likelihood of the muon hypothesis, 
as provided by the muon system, with respect to the~pion hypothesis, 
$\Delta^{\Pmu/\Ppi} \ln \mathcal{L}$~\cite{Muon:performance},
is greater than zero.  Good quality particle identification 
by the ring-imaging Cherenkov detectors is ensured by requiring 
the momentum of  the~hadron 
candidates, $p$, 
to be between $3.2\gevc$ and $100\gevc$, and  the 
pseudorapidity to be in the range $2<\eta<5$. 
To select well-identified kaons~(pions) the corresponding 
difference in logarithms of  the likelihood of the kaon and pion 
hypotheses~\cite{arXiv:1211-6759}
is required to be 
$\Delta^{\mathrm{K}/\Ppi} \ln \mathcal{L}>2(<0)$. 
These criteria are chosen
to be tight enough to reduce significantly 
the background due to misidentification,
whilst ensuring good agreement between data and simulation. 

To ensure that the hadrons used in the analysis are inconsistent with 
being directly produced in a  pp~interaction vertex, 
the impact parameter $\chi^2$, defined as the difference 
between the $\chi^2$ of 
the reconstructed pp~collision vertex 
formed with and without the considered track, 
is required to be $\chi^2_{\mathrm{IP}}>9$.
When more than one vertex is reconstructed, 
that with the smallest value 
of $\chi^2_{\mathrm{IP}}$~is chosen.

As in Refs.~\cite{LHCb-PAPER-2012-022,LHCb-PAPER-2012-010,LHCb-PAPER-2012-053} 
the selection of $\jpsimm$ candidates proceeds from pairs of 
oppositely-charged muons  forming a common vertex. 
The quality of the vertex is ensured 
by requiring the $\chi^2$ of the vertex fit, $\chi^2_{\mathrm{vx}}$, 
to be less  than~30.
The vertex is forced to be well separated 
from the reconstructed pp~interaction vertex by requiring 
the decay length significance,
$\mathcal{S}_{\mathrm{flight}}$, defined as 
the ratio of the projected distance 
from pp~interaction vertex to \mumu~vertex on 
direction of \mumu~pair momentum
and its uncertainty,
to be greater than 3. 
Finally, the mass of the dimuon combination is required to be within
$\pm45\mevcc$~of the known \jpsi~mass~\cite{PDG2012},
which corresponds to a $\pm3.5\sigma$ window, where 
$\sigma$~is the measured \jpsi~mass resolution. 

Candidate \Ds~mesons are reconstructed in the 
\dsphipi~mode using criteria 
similar to those in Ref.~\cite{LHCb-PAPER-2012-003}. 
A  good vertex quality is ensured by requiring $\chi^2_{\mathrm{vx}}<25$.
The mass of the kaon pair is required to be consistent 
with the decay $\Pphi{}\to\Kp{}\Km$,
\mbox{$\left| m_{\Kp{}\Km} - m_{\Pphi} \right| < 20\mevcc$}. 
Finally, the
mass of the candidate is required to be within 
$\pm20\mevcc$~of the~known \Ds~mass~\cite{PDG2012},
which corresponds to a $\pm3.5\sigma$ window, where 
$\sigma$~is the measured \Ds~mass resolution,   
and its transverse momentum to be $>1\gevc$. 

Candidate \Bc~mesons are formed from $\jpsi{}\Ds$~pairs
with transverse momentum in excess of 1\gevc.
The candidates should be consistent with being 
produced in a~$\mathrm{pp}$~interaction vertex by 
requiring $\chi^2_{\mathrm{IP}}<9$ with 
respect to reconstructed 
pp~collision vertices.
A~kinematic fit is applied to the~\Bc~candidates~\cite{Hulsbergen:2005pu}. 
To improve the mass and lifetime resolution, in this fit, 
a~constraint on the pointing of the candidate to the primary vertex is applied
together with mass constraints on the intermediate 
\jpsi~and \Ds~states. The value of the \jpsi~mass is taken from
Ref.~\cite{PDG2012}. 
For the \Ds meson the
value of $m_{\Ds} = 1968.31\pm0.20\mevcc$ 
is used, that is the average of the values given 
in  Ref.~\cite{LHCb-PAPER-2013-011} 
and Ref.~\cite{PDG2012}. 
The $\chi^2$ per degree of freedom of this fit, 
$\chi^2_{\mathrm{fit}}/\mathrm{ndf}$, is required to be less than $5$. 
The decay time of the \Ds~candidate, $c\tau\left(\Ds\right)$, 
determined by this fit, 
is required to satisfy $c\tau>75\mum$.
The corresponding signed 
significance, $\mathcal{S}_{c\tau}$, 
defined as  the ratio of  
the measured decay time  
and its uncertainty,
 is required to be 
in excess of~$3$. 
Finally, the decay time
of the \Bc~candidate, $c\tau\left(\Bc\right)$, 
is required to be between~$75\mum$ and~1\mm.
The upper edge, in excess of 7~lifetimes of \Bc~meson,
is introduced to remove badly recontructed candidates.

\section{Observation of {\boldmath\bctods}}    
\label{seq:btods}
\begin{figure}[t]
  \setlength{\unitlength}{1mm}
  \centering
  \includegraphics*[width=150mm,height=120mm%
  ]{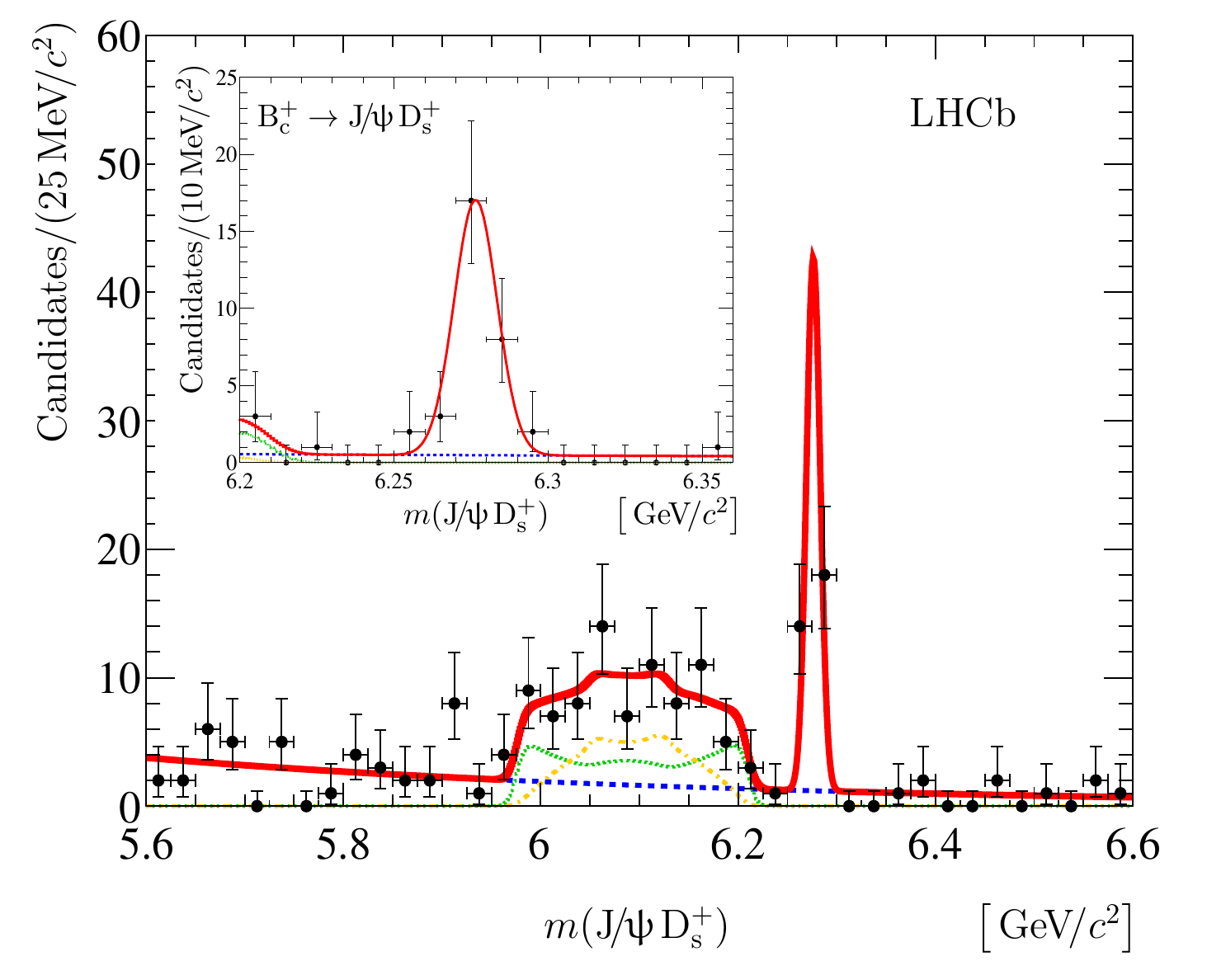}
  \caption { \small
    Mass distributions for selected \jpsi{}\Ds~pairs. 
    The solid curve represents the result of a fit to the model
    described in the text. 
    The contribution from 
    the $\Bc\to\jpsi\D^{\ast+}_{\mathrm{s}}$~decay 
    is shown with thin green dotted 
    and thin yellow dash-dotted lines
    for the $\mathcal{A}_{\pm\pm}$ and $\mathcal{A}_{00}$ amplitudes,
    respectively. 
    The insert shows a zoom of the \Bc~mass region.
  }
  \label{fig:fits}
\end{figure}

The mass distribution of the selected \bctods~candidates is
shown in Fig.~\ref{fig:fits}.
The~peak close to the known mass of 
the \Bc~meson~\cite{PDG2012,LHCb-PAPER-2012-028} 
with a width 
compatible with the~expected mass resolution is interpreted as being 
due to the \bctods~decay.
The wide structure between 5.9~and $6.2\gevcc$ is 
attributed to the decay  $\Bc{}\to{}\jpsi{}\D^{\ast+}_{\mathrm{s}}$, 
followed by $\D^{\ast+}_{\mathrm{s}}\to{}\Ds{}\Pgamma$ or 
$\D^{\ast+}_{\mathrm{s}}\to{}\Ds{}\Ppi^0$~decays, where the neutral particles
are not detected.  
The~process $\Bc{}\to{}\jpsi{}\D^{\ast+}_{\mathrm{s}}$ being the decay 
of a~pseudoscalar particle into two vector particles 
is described  by three helicity amplitudes:
$\mathcal{A}_{++}$,
$\mathcal{A}_{00}$ and 
$\mathcal{A}_{--}$, where indices 
correspond to the helicities 
of the \jpsi~and $\D^{\ast+}_{\mathrm{s}}$ mesons.
Simulation studies show that the  \jpsi{}\Ds 
mass distributions are the same for 
the $\mathcal{A}_{++}$ and $\mathcal{A}_{--}$ amplitudes. 
Thus,  the \jpsi{}\Ds~mass spectrum is described 
by a~model consisting of the following components:
an~exponential shape to describe the combinatorial 
background, a~Gaussian shape to describe 
the~\mbox{\bctods}~signal
and two helicity components to describe the
\mbox{$\Bc\to\jpsi\D^{\ast+}_{\mathrm{s}}$}~contributions
corresponding to the $\mathcal{A}_{\pm\pm}$ and 
                     $\mathcal{A}_{00}$ amplitudes.
The shape of
these components is determined using the simulation where  
the branching fractions for $\D^{\ast+}_{\mathrm{s}}\to \Ds \Pgamma$ and 
$\D^{\ast+}_{\mathrm{s}}\to \Ds \Ppi^0$ decays are 
taken from Ref.~\cite{PDG2012}.

To estimate the signal yields, an extended unbinned maximum likelihood fit to
the mass distribution is performed. The correctness of the fit 
procedure together with the reliability of the estimated uncertainties
has been extensively checked using simulation.
The fit has seven free parameters: the mass of the \Bc~meson, $m_{\Bc}$, 
the signal resolution, $\sigma_{\Bc}$, 
the relative amount 
of the $\mathcal{A}_{\pm\pm}$~helicity 
amplitudes 
of total \bctodsst~decay rate, 
$\mathrm{f}_{\pm\pm}$, 
the slope parameter of the exponential
background and the yields of the two signal components,
$N_{\bctods}$ and $N_{\bctodsst}$, and of the background. 
The values of the~signal parameters obtained from the fit
are summarized in Table~\ref{tab:fit_results}. 
The fit result is also shown in Fig.~\ref{fig:fits}.

\begin{table}[t]
\centering
\caption{\small 
  Signal parameters of the unbinned 
  extended maximum likelihood
  fit to the \jpsi{}\Ds~mass distribution.}
\vspace*{3mm}
\begin{tabular*}{0.9\textwidth}{@{\hspace{15mm}}lc@{\extracolsep{\fill}}c@{\hspace{15mm}}}
  \multicolumn{2}{c}{Parameter}  & Value \\ \hline  
  $m_{\Bc} $                        &  $\left[\mevcc\right]$ & $6276.28 \pm 1.44\phantom{000}$   \\
  $\sigma_{\Bc}                   $ &  $\left[\mevcc\right]$ & $\phantom{0}7.0     \pm 1.1\phantom{0}$     \\ 
  $N_{\bctods}                     $ &                        & $28.9    \pm 5.6\phantom{0}$     \\ 
  $\dfrac{N_{\bctodsst}}{ N_{\bctods} }$ &                       & $2.37    \pm 0.56$ \\
  $\mathrm{f}_{\pm\pm} $             &  $\left[\%\right]$     & $52      \pm 20$  
\end{tabular*}
\label{tab:fit_results}
\end{table}

To check the result, the fit has been performed with different models 
for the signal: a~double-sided Crystal Ball 
function~\cite{Skwarnicki:1986xj,LHCb-PAPER-2011-013}, 
and a modified Novosibirsk function~\cite{Lees:2011gw}. 
For~these tests the tail and asymmetry parameters are fixed using 
the simulation values, while the parameters representing the 
peak position and resolution are left free to vary. 
As~alternative models for
the background, the product of an exponential function and 
a~fourth-order polynomial  function are used. 
The fit parameters obtained 
are  stable with respect to the~choice of the fit model 
and the fit range
interval.

The statistical significance for the \bctods~signal is estimated 
from the change in the likelihood function 
$\mathcal{S}_{\sigma} = \sqrt{ 2 \ln 
\tfrac { \mathcal{L}_{\mathcal{B}+\mathcal{S}}}{ \mathcal{L}_{\mathcal{B}}}}$,
where
$\mathcal{L}_{\mathcal{B}}$ is the likelihood of a background-only hypothesis and 
$\mathcal{L}_{\mathcal{B}+\mathcal{S}}$ is the likelihood of a background-plus-signal 
hypothesis.  The~significance has been estimated separately 
for the \bctods and \BctoDsst~signals.
To~exclude the look-elsewhere effect~\cite{Lyons:2008,*Gross:2010}, 
the mass and resolution
of the peak are fixed to the values obtained with the simulation. 
The~minimal significance found varying the fit
model as described above is taken as the signal significance. 
The statistical significance for both the \bctods and
\BctoDsst~signals estimated in this way is in excess of 
9~standard deviations.

The low $Q$-value for the \bctods~decay mode allows 
the \Bc~mass to be precisely measured. 
This makes use of the \Ds~mass value, 
evaluated in Sect.~\ref{seq:evsel}, 
taking correctly into account the correlations between 
the measurements. 
The calibration of the momentum scale for the
dataset used here is detailed in Refs.~\cite{LHCb-PAPER-2012-048,LHCb-PAPER-2013-011}. 
It is based upon large calibration samples of $\Bp{}\to{}\jpsi{}\Kp$ and
$\jpsi\to\mumu$ decays and leads to an~accuracy in the momentum scale 
of $3\times10^{-4}$. This translates into an
uncertainty of 0.30\mevcc on the \Bc~meson mass. A further uncertainty of   
0.11\mevcc arises from the knowledge of the detector material 
distribution~\cite{LHCb-PAPER-2012-048,LHCb-PAPER-2013-011,LHCb-PAPER-2012-028,LHCb-PAPER-2011-035} 
and the signal modelling. The uncertainty on the \Ds~mass
results in a 0.16\mevcc uncertainty on the \Bc~meson mass. Adding
these in quadrature gives
\begin{equation*}
m_{\Bc} =  6276.28 \pm 1.44\,\stat \pm 0.36\,\syst\mevcc.
\end{equation*}  
The uncertainty on the \Ds~meson mass and on the momentum scale
largely cancels in the mass difference
\begin{equation*}
m_{\Bc} - m_{\Ds}  =  4307.97 \pm 1.44\,\stat \pm 0.20\,\syst\mevcc.
\end{equation*}

\section{Normalization to the {\boldmath\bctop} decay mode}
\label{seq:norm} 
\begin{figure}[htb]
  \setlength{\unitlength}{1mm}
  \centering
  \includegraphics*[width=150mm,height=120mm%
  ]{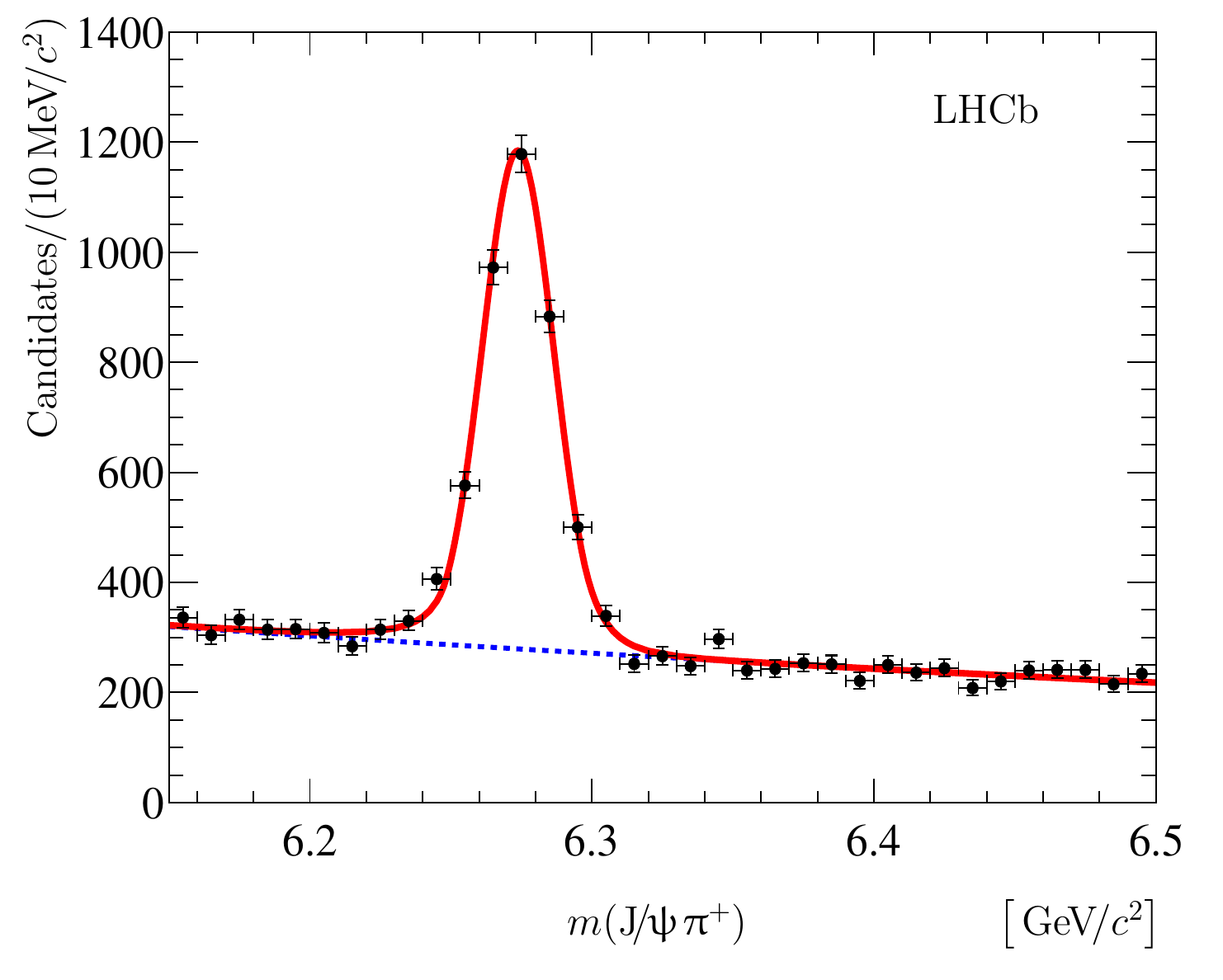}
  \caption { \small
    Mass distribution for selected \bctop~candidates. The results of
    a fit to the model described in the text are superimposed (solid
    line) together with the background component (dotted line).
  }
  \label{fig:norm}
\end{figure}

A large sample of \bctop~decays
serves as a~normalization channel
to measure the~ratio of branching fractions 
for the \bctods~and \bctop~modes. 
Selection of \bctop~events is performed in a manner similar
to that described in
Sect.~\ref{seq:evsel} for the~signal channel. 
To further reduce the combinatorial background,
the~transverse momentum of the pion
for the \bctop~mode is required to be in excess of 1\gevc.
The mass distribution of the selected \bctop~candidates is shown 
in Fig.~\ref{fig:norm}.

To determine the yield, an extended unbinned maximum likelihood
fit to the mass distribution is performed. 
The signal is modelled by a double-sided Crystal Ball 
function and the background with an exponential function. 
The fit gives a yield of $3009 \pm 79$ events.
As cross-checks, a modified Novosibirsk function
and a~Gaussian function for the signal component and a 
product of exponential and polynomial functions for the background are
used. The difference is treated as systematic uncertainty.

The ratio of the total efficiencies (including acceptance, reconstruction, 
selection and trigger)
for the \bctods and \bctop modes is determined with  
simulated data to be $0.148\pm0.001$, where the uncertainty 
is statistical only.  As only events explicitly selected 
by the \jpsi~triggers are used, 
the~ratio of the trigger efficiencies for 
the \bctods~and \bctop~modes is close to unity.


\section{Systematic uncertainties}
\label{seq:syst}  

\begin{figure}[t]
  \setlength{\unitlength}{1mm}
  \centering
  \includegraphics*[width=150mm,height=60mm%
  ]{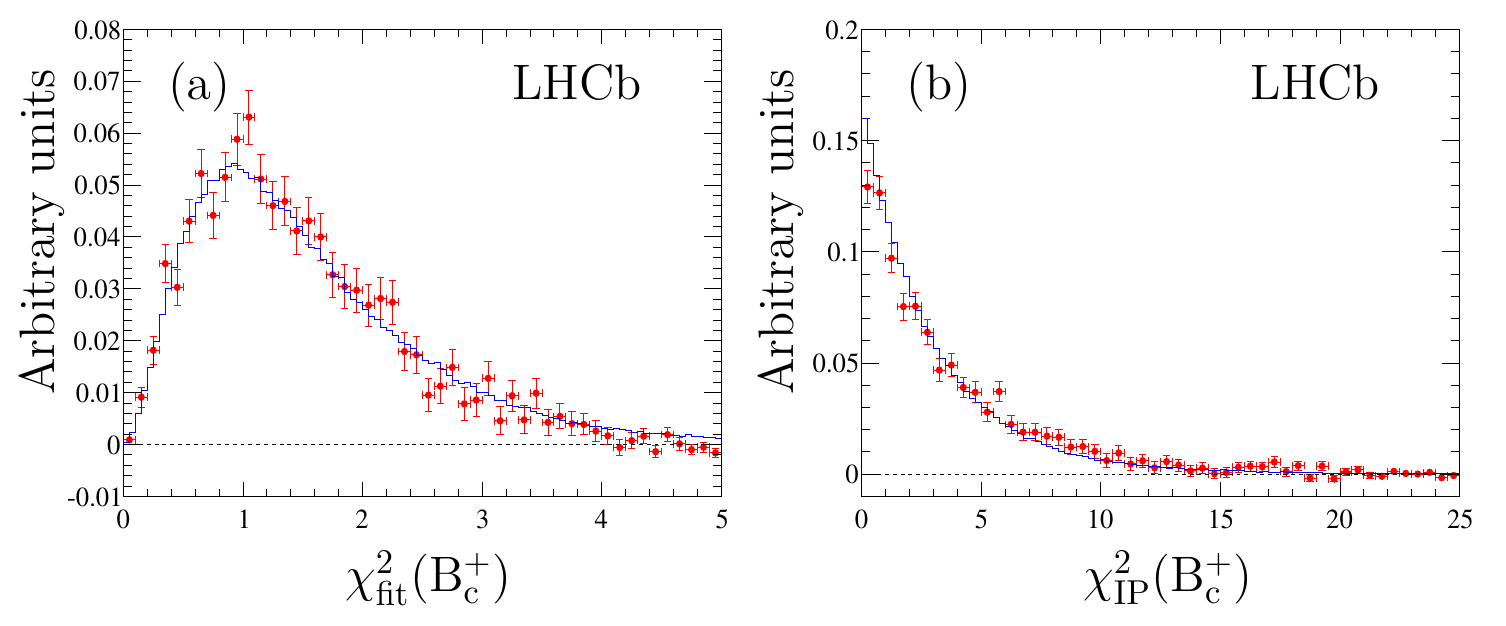}
  \caption { \small
    Distributions of 
    (a)~$\chi^2_{\mathrm{fit}}(\Bc)$ and 
    (b)~$\chi^2_{\mathrm{IP}}(\Bc)$
    for \bctop~events: background subtracted data  
    (red points with error bars),
    and simulation (blue histogram).
  }
  \label{fig:sys_c2dtf}
\end{figure}

Uncertainties on the ratio~\rdspi \  
related to differences between the data and simulation 
efficiency 
for the~selection requirements 
are studied using the abundant \bctop~channel.
As an example, Fig.~\ref{fig:sys_c2dtf} compares the distributions of 
 $\chi^2_{\mathrm{fit}}(\Bc)$ and 
 $\chi^2_{\mathrm{IP}}(\Bc)$ 
for data and simulated \bctop~events. 
For background subtraction 
the \sPlot~techinque~\cite{Pivk:2004ty} has been used.
It can be seen that the agreement between data and
simulation is good. In addition, a large sample of selected 
$\Bp{}\to{}\jpsi{}\left(\Kp{}\Km\right)_{\Pphi}\Kp$~events
has been used to quantify differences between data and simulation. 
Based on the deviation, a systematic uncertainty of~1\% is assigned.

The agreement of the absolute trigger efficiency between data and
simulation has been validated to a precision of 4\% 
using the technique described in 
Refs.~\cite{LHCb-PAPER-2010-001,LHCb-PAPER-2011-013,Aaij:2012me} 
with a large sample of $\Bp{}\to{}\jpsi{}\left(\Kp{}\Km\right)_{\Pphi}\Kp$~events.
A~further cancellation of uncertainties occurs in the~ratio of 
branching fractions resulting in a systematic uncertainty of $1.1\%$.

The systematic uncertainties related to the fit model, in particular to the 
signal shape, mass and resolution for the \bctods~mode and the fit
interval have been discussed in Sects.~\ref{seq:btods} and~\ref{seq:norm}. 
The main part comes from the normalization channel~\bctop.

Other systematic uncertainties arise from differences in the
efficiency of charged particle reconstruction between data and simulation. 
The largest of these arises from the knowledge of the
hadronic interaction probability in the detector,
which has an uncertainty of 
$2\%$~per track~\cite{LHCb-PAPER-2010-001}. A further uncertainty related 
to the reconstruction of two additional kaons in the \bctods~mode 
with respect to the \bctop~mode 
is estimated to be 
$2\times0.6\%$~\cite{LHCb-PUB-2011-025}.
Further uncertainties are related 
to the track quality selection requirements $\chi^2_{\mathrm{tr}}<4$ and 
$\mathcal{P}_{\mathrm{fake}}<0.5$. These are estimated from a comparison
of data and simulation in the \bctop~decay mode 
to be $0.4\%$ per final state track.

The uncertainty associated with the kaon identification 
criteria is studied using the~combined \bctods~and \BctoDsst~signals. 
The efficiency to identify a kaon pair 
with a selection on $\Delta^{\mathrm{K}/\Ppi}\ln \mathcal{L} $ 
has been compared for data and simulation for various selection requirements. 
The comparison shows a  $(-1.8\pm2.9)\%$ difference between 
data and simulation in the efficiency to identify 
a kaon pair with $2 \le \min \Delta^{\mathrm{K}/\Ppi} \log\mathcal{L}$.
This estimate has been confirmed using a~kinematically similar sample of 
reconstructed
$\Bp{}\to{}\jpsi{}\left(\Kp{}\Km\right)_{\Pphi}\Kp$~events. 
An uncertainty of 3\% is assigned.

The limited knowledge of the \Bc~lifetime leads to 
an additional systematic uncertainty due to the different decay 
time acceptance between the \bctods~and 
\bctop~decay modes. To estimate this effect, 
the decay time distributions for simulated events are reweighted to change 
the \Bc~lifetime by one standard deviation 
from the known value~\cite{PDG2012}, as well as the value 
recently measured by the CDF collaboration~\cite{Aaltonen:2012yb},   
and the efficiencies are recomputed. An uncertainty of $1\%$ is assigned.

Possible uncertainties related to the stability of the data taking
conditions are tested by studying the ratio of 
the yields of $\Bp{}\to{}\jpsi{}\Kp{}\pipi$ and 
$\Bp{}\to{}\jpsi{}\Kp$~decays for different data taking periods 
and dipole magnet polarities. 
This results in a further $2.5\%$ uncertainty.

The largest systematic uncertainty is due to the knowledge of the  branching fraction 
of the $\Ds{}\to{}\left(\Km{}\Kp{}\right)_{\Pphi}{}\pip$~decay, 
with a kaon pair mass within $\pm20\mevcc$ of the known 
\Pphi~meson mass. The value of  $(2.24\pm0.11\pm0.06)\%$ from
Ref.\cite{Alexander:2008aa}  is used in the analysis.
The systematic uncertainties
on \rdspi~are summarized in Table~\ref{tab:syst}.

\begin{table}[htb]
  \centering
  \caption{ \small
    Relative systematic uncertainties 
    for the ratio of branching fractions 
    of \mbox{\bctods} and \mbox{\bctop}.
  } \label{tab:syst}
  \vspace*{3mm}
  \begin{tabular*}{0.8\textwidth}{@{\hspace{10mm}}l@{\extracolsep{\fill}}c@{\hspace{10mm}}}
    Source & Uncertainty~$\left[\%\right]$  
    \\
    \hline 
    Simulated efficiencies            &  1.0  
    \\
    Trigger                           &  1.1 
    \\
    Fit model                         &  1.8
    \\ 
    Track reconstruction              &  $2\times0.6$
    \\
    Hadron interactions               &  $2\times2.0$
    \\ 
    Track quality selection           &  $2\times0.4$
    \\
    Kaon  identification              &  3.0 
    \\
    \Bc~lifetime                      &  1.0 
    \\ 
    Stability for various data taking conditions & 2.5 
    \\
    $\BR\left(\Ds\to\left(\Km\Kp\right)_{\Pphi}\pip\right)$ &  5.6 
    \\
    \hline 
    Total    &  8.4 
  \end{tabular*}   
\end{table}

The ratio \rdsds~is estimated  as 
\begin{equation}
\rdsds =  \frac{N_{\bctodsst}}{ N_{\bctods} },
\end{equation}
where the ratio of yields is given in Table~\ref{tab:fit_results}.
The uncertainty associated with the assumption that the efficiencies
for the \bctods and \bctodsst modes are equal, is evaluated by studying
the dependence of the relative yields for these modes for 
loose (or no) requirements on  the $\chi^2_{\mathrm{IP}}(\Bc)$,  
$\chi^2_{\mathrm{fit}}(\Bc)$ and $c\tau({}\Bc)$~variables. 
For this selection the measured ratio of 
\mbox{\BctoDsst} to \mbox{\bctods}~events changes to $2.27\pm0.59$. 
An uncertainty of 4\%  
is assigned to the \rdsds~ratio.

The uncertainty on the fraction of the $\mathcal{A}_{\pm\pm}$
amplitude, $\mathrm{f}_{\pm\pm}$, has been studied with 
different fit models for the parameterization of 
the combinatorial background, as well 
as different mass resolution models. This is negligible in comparison 
to the statistical uncertainty.

\section{Results and summary}

The decays \bctods and \BctoDsst have been observed 
for the first time with 
statistical significances in excess of 9~standard deviations. 
The ratio of branching fractions for \bctods and \bctop is calculated as 
\begin{equation}
  \dfrac {\BR \left( \bctods \right) } 
         {\BR \left( \bctop  \right) } 
      = 
  \dfrac  {1}
          {\BR_{\Ds}}
  \times 
  \dfrac  {\varepsilon^{\mathrm{tot}}_{\bctop}}
          {\varepsilon^{\mathrm{tot}}_{\bctods}}
  \times 
  \dfrac  {N \left( \bctods \right) } 
          {N \left( \bctop  \right) },
\end{equation}
  where the value 
  of $\BR_{\Ds} = \BR\left(\Ds{}\to{}\left(\Km{}\Kp{}\right)_{\Pphi}{}\pip\right)$~\cite{Alexander:2008aa}  
  with the mass of the
  kaon pair within $\pm 20\mevcc$ of the known value 
  of the $\Pphi$
  mass is used, 
  together with
  the ratio of efficiencies,
  and the signal yields given in Sects.~\ref{seq:btods} and~\ref{seq:norm}. 
  This results in
\begin{equation*}
  \dfrac {\BR \left( \bctods \right) } 
         {\BR \left( \bctop  \right) } = 
      2.90 \pm 0.57\,\stat \pm 0.24\,\syst.\label{eq:results}
\end{equation*}
The value obtained is in agreement 
with the na\"ive expectations given in Eq.~\eqref{eq:exp_fact}
from \Bz~decays,  and the values from Refs.~\cite{Colangelo:1999zn,Dhir:2008hh,Ivanov:2006ni} 
but larger than predictions from Refs.~\cite{Kiselev:2003mp,Chang:1992pt}
and factorization expectations from \Bp~decays. 

The ratio of branching fractions for 
the \BctoDsst~and \bctods~decays 
is measured to be 
\begin{equation*}
\dfrac
{  \BR\left( \BctoDsst \right) }
{  \BR\left( \BctoDs   \right) } = 2.37 \pm 0.56\,\stat \pm 0.10\,\syst.
\end{equation*}
This result is in agreement with the na\"ive
factorization hypothesis (Eq.~\eqref{eq:exp_fact2}) 
and with the predictions of 
Refs.~\cite{Ivanov:2006ni, Colangelo:1999zn}.

The fraction of the $\mathcal{A}_{\pm\pm}$
amplitude in the \BctoDsst~decay is measured to be  
\begin{equation*}
\dfrac
{  \Gamma_{\pm\pm}     \left( \BctoDsst \right) }
{  \Gamma_{\mathrm{tot}} \left( \BctoDsst \right) }
= (52\pm20)\%,
\end{equation*}
in agreement with a simple estimate of $\tfrac{2}{3}$,
the measurements~\cite{Ahmed:2000ad,Aubert:2003jj} and 
factorization predictions~\cite{Richman}
for  $\Bd\to\D^{*-}\D_{\squark}^{\ast+}$~decays,
and expectations for 
$\Bc\to\jpsi\ell^{+}\Pnu_{\ell}$~decays
from Refs.~\cite{Ebert:2003cn,Likhoded:2009ib}.

The mass of the \Bc~meson and the mass difference between the
\Bc~and \Ds~mesons are measured to be
\begin{eqnarray*}
m_{\Bc}          & = & 6276.28 \pm 1.44\,\stat \pm 0.36\,\syst\mevcc, \\  
m_{\Bc} - m_{\Ds} & = & 4307.97 \pm 1.44\,\stat \pm 0.20\,\syst\mevcc.
\end{eqnarray*} 
The \Bc~mass measurement is in good agreement with the
previous result obtained by LHCb in the \bctop 
mode~\cite{LHCb-PAPER-2012-028} and has smaller 
systematic uncertainty.

\section*{Acknowledgements}

\noindent 
We thank A.~Luchinsky and A.K.~Likhoded
for advice on aspects of \Bc physics.
%
We express our gratitude to our colleagues in the CERN
accelerator departments for the excellent performance of the LHC. We
thank the technical and administrative staff at the LHCb
institutes. We acknowledge support from CERN and from the national
agencies: CAPES, CNPq, FAPERJ and FINEP (Brazil); NSFC (China);
CNRS/IN2P3 and Region Auvergne (France); BMBF, DFG, HGF and MPG
(Germany); SFI (Ireland); INFN (Italy); FOM and NWO (The Netherlands);
SCSR (Poland); ANCS/IFA (Romania); MinES, Rosatom, RFBR and NRC
``Kurchatov Institute'' (Russia); MinECo, XuntaGal and GENCAT (Spain);
SNSF and SER (Switzerland); NAS Ukraine (Ukraine); STFC (United
Kingdom); NSF (USA). We also acknowledge the support received from the
ERC under FP7. The Tier1 computing centres are supported by IN2P3
(France), KIT and BMBF (Germany), INFN (Italy), NWO and SURF (The
Netherlands), PIC (Spain), GridPP (United Kingdom). We are thankful
for the computing resources put at our disposal by Yandex LLC
(Russia), as well as to the communities behind the multiple open
source software packages that we depend on.

\addcontentsline{toc}{section}{References}

\bibliographystyle{LHCb}
\bibliography{main,LHCb-PAPER,LHCb-CONF,local}

\ifx\mcitethebibliography\mciteundefinedmacro
\PackageError{LHCb.bst}{mciteplus.sty has not been loaded}
{This bibstyle requires the use of the mciteplus package.}\fi
\providecommand{\href}[2]{#2}
\begin{mcitethebibliography}{10}
\mciteSetBstSublistMode{n}
\mciteSetBstMaxWidthForm{subitem}{\alph{mcitesubitemcount})}
\mciteSetBstSublistLabelBeginEnd{\mcitemaxwidthsubitemform\space}
{\relax}{\relax}

\bibitem{PDG2012}
Particle Data Group, J.~Beringer {\em et~al.},
  \ifthenelse{\boolean{articletitles}}{{\it {\href{http://pdg.lbl.gov/}{Review
  of particle physics}}},
  }{}\href{http://dx.doi.org/10.1103/PhysRevD.86.010001}{Phys.\ Rev.\  {\bf
  D86} (2012) 010001}\relax
\mciteBstWouldAddEndPuncttrue
\mciteSetBstMidEndSepPunct{\mcitedefaultmidpunct}
{\mcitedefaultendpunct}{\mcitedefaultseppunct}\relax
\EndOfBibitem
\bibitem{Aaltonen:2012yb}
CDF collaboration, T.~Aaltonen {\em et~al.},
  \ifthenelse{\boolean{articletitles}}{{\it {Measurement of the
  $\B_{\cquark}^{-}$~meson lifetime in the decay $\B_{\cquark}^{-} \rightarrow
  \jpsi~\pim$}}, }{}\href{http://dx.doi.org/10.1103/PhysRevD.87.011101}{Phys.\
  Rev.\  {\bf D87} (2013) 011101}, \href{http://arxiv.org/abs/1210.2366}{{\tt
  arXiv:1210.2366}}\relax
\mciteBstWouldAddEndPuncttrue
\mciteSetBstMidEndSepPunct{\mcitedefaultmidpunct}
{\mcitedefaultendpunct}{\mcitedefaultseppunct}\relax
\EndOfBibitem
\bibitem{Abe:1998wi}
CDF collaboration, F.~Abe {\em et~al.},
  \ifthenelse{\boolean{articletitles}}{{\it {Observation of the \Bc~meson in
  $\mathrm{p}\bar{\mathrm{p}}$~collisions at $\sqrt{s} = 1.8\tev$}},
  }{}\href{http://dx.doi.org/10.1103/PhysRevLett.81.2432}{Phys.\ Rev.\ Lett.\
  {\bf 81} (1998) 2432}, \href{http://arxiv.org/abs/hep-ex/9805034}{{\tt
  arXiv:hep-ex/9805034}}\relax
\mciteBstWouldAddEndPuncttrue
\mciteSetBstMidEndSepPunct{\mcitedefaultmidpunct}
{\mcitedefaultendpunct}{\mcitedefaultseppunct}\relax
\EndOfBibitem
\bibitem{Abulencia:2005usa}
CDF collaboration, A.~Abulencia {\em et~al.},
  \ifthenelse{\boolean{articletitles}}{{\it {Evidence for the exclusive decay
  $\B_{\cquark}^{\pm} \to \jpsi \Ppi^\pm$ and measurement of the mass of the
  \Bc~meson}}, }{}\href{http://dx.doi.org/10.1103/PhysRevLett.96.082002}{Phys.\
  Rev.\ Lett.\  {\bf 96} (2006) 082002},
  \href{http://arxiv.org/abs/hep-ex/0505076}{{\tt arXiv:hep-ex/0505076}}\relax
\mciteBstWouldAddEndPuncttrue
\mciteSetBstMidEndSepPunct{\mcitedefaultmidpunct}
{\mcitedefaultendpunct}{\mcitedefaultseppunct}\relax
\EndOfBibitem
\bibitem{LHCb-PAPER-2011-044}
LHCb collaboration, R.~Aaij {\em et~al.},
  \ifthenelse{\boolean{articletitles}}{{\it {First observation of the decay
  $\Bc \to \jpsi \pi^+\pi^-\pi^+$}},
  }{}\href{http://dx.doi.org/10.1103/PhysRevLett.108.251802}{Phys.\ Rev.\
  Lett.\  {\bf 108} (2012) 251802}, \href{http://arxiv.org/abs/1204.0079}{{\tt
  arXiv:1204.0079}}\relax
\mciteBstWouldAddEndPuncttrue
\mciteSetBstMidEndSepPunct{\mcitedefaultmidpunct}
{\mcitedefaultendpunct}{\mcitedefaultseppunct}\relax
\EndOfBibitem
\bibitem{LHCb-PAPER-2012-054}
LHCb collaboration, R.~Aaij {\em et~al.},
  \ifthenelse{\boolean{articletitles}}{{\it {Observation of
  $\Bc\to\psitwos\pip$}}, }{}\href{http://arxiv.org/abs/1303.1737}{{\tt
  arXiv:1303.1737}}, {submitted to Phys. Rev. D}\relax
\mciteBstWouldAddEndPuncttrue
\mciteSetBstMidEndSepPunct{\mcitedefaultmidpunct}
{\mcitedefaultendpunct}{\mcitedefaultseppunct}\relax
\EndOfBibitem
\bibitem{Kiselev:2003mp}
V.~Kiselev, \ifthenelse{\boolean{articletitles}}{{\it {Decays of the
  \Bc~meson}}, }{}\href{http://arxiv.org/abs/hep-ph/0308214}{{\tt
  arXiv:hep-ph/0308214}}\relax
\mciteBstWouldAddEndPuncttrue
\mciteSetBstMidEndSepPunct{\mcitedefaultmidpunct}
{\mcitedefaultendpunct}{\mcitedefaultseppunct}\relax
\EndOfBibitem
\bibitem{Colangelo:1999zn}
P.~Colangelo and F.~De~Fazio, \ifthenelse{\boolean{articletitles}}{{\it {Using
  heavy quark spin symmetry in semileptonic \Bc~decays}},
  }{}\href{http://dx.doi.org/10.1103/PhysRevD.61.034012}{Phys.\ Rev.\  {\bf
  D61} (2000) 034012}, \href{http://arxiv.org/abs/hep-ph/9909423}{{\tt
  arXiv:hep-ph/9909423}}\relax
\mciteBstWouldAddEndPuncttrue
\mciteSetBstMidEndSepPunct{\mcitedefaultmidpunct}
{\mcitedefaultendpunct}{\mcitedefaultseppunct}\relax
\EndOfBibitem
\bibitem{Ivanov:2006ni}
M.~A. Ivanov, J.~G. Korner, and P.~Santorelli,
  \ifthenelse{\boolean{articletitles}}{{\it {Exclusive semileptonic and
  nonleptonic decays of the \Bc~meson}},
  }{}\href{http://dx.doi.org/10.1103/PhysRevD.73.054024}{Phys.\ Rev.\  {\bf
  D73} (2006) 054024}, \href{http://arxiv.org/abs/hep-ph/0602050}{{\tt
  arXiv:hep-ph/0602050}}\relax
\mciteBstWouldAddEndPuncttrue
\mciteSetBstMidEndSepPunct{\mcitedefaultmidpunct}
{\mcitedefaultendpunct}{\mcitedefaultseppunct}\relax
\EndOfBibitem
\bibitem{Dhir:2008hh}
R.~Dhir and R.~Verma, \ifthenelse{\boolean{articletitles}}{{\it {\Bc~meson
  form-factors and $\Bc \to \mathrm{PV}$ decays involving flavor dependence of
  transverse quark momentum}},
  }{}\href{http://dx.doi.org/10.1103/PhysRevD.79.034004}{Phys.\ Rev.\  {\bf
  D79} (2009) 034004}, \href{http://arxiv.org/abs/0810.4284}{{\tt
  arXiv:0810.4284}}\relax
\mciteBstWouldAddEndPuncttrue
\mciteSetBstMidEndSepPunct{\mcitedefaultmidpunct}
{\mcitedefaultendpunct}{\mcitedefaultseppunct}\relax
\EndOfBibitem
\bibitem{Chang:1992pt}
C.-H. Chang and Y.-Q. Chen, \ifthenelse{\boolean{articletitles}}{{\it {The
  decays of \Bc~meson}},
  }{}\href{http://dx.doi.org/10.1103/PhysRevD.49.3399}{Phys.\ Rev.\  {\bf D49}
  (1994) 3399}\relax
\mciteBstWouldAddEndPuncttrue
\mciteSetBstMidEndSepPunct{\mcitedefaultmidpunct}
{\mcitedefaultendpunct}{\mcitedefaultseppunct}\relax
\EndOfBibitem
\bibitem{Alves:2008zz}
LHCb collaboration, A.~A. Alves~Jr. {\em et~al.},
  \ifthenelse{\boolean{articletitles}}{{\it {The \lhcb detector at the LHC}},
  }{}\href{http://dx.doi.org/10.1088/1748-0221/3/08/S08005}{JINST {\bf 3}
  (2008) S08005}\relax
\mciteBstWouldAddEndPuncttrue
\mciteSetBstMidEndSepPunct{\mcitedefaultmidpunct}
{\mcitedefaultendpunct}{\mcitedefaultseppunct}\relax
\EndOfBibitem
\bibitem{Aaij:2012me}
R.~Aaij {\em et~al.}, \ifthenelse{\boolean{articletitles}}{{\it {The LHCb
  Trigger and its Performance in 2011}},
  }{}\href{http://dx.doi.org/10.1088/1748-0221/8/04/P04022}{JINST {\bf 8}
  (2013) P04022}, \href{http://arxiv.org/abs/1211.3055}{{\tt
  arXiv:1211.3055}}\relax
\mciteBstWouldAddEndPuncttrue
\mciteSetBstMidEndSepPunct{\mcitedefaultmidpunct}
{\mcitedefaultendpunct}{\mcitedefaultseppunct}\relax
\EndOfBibitem
\bibitem{Sjostrand:2006za}
T.~Sj\"{o}strand, S.~Mrenna, and P.~Skands,
  \ifthenelse{\boolean{articletitles}}{{\it {\pythia~6.4 physics and manual}},
  }{}\href{http://dx.doi.org/10.1088/1126-6708/2006/05/026}{JHEP {\bf 05}
  (2006) 026}, \href{http://arxiv.org/abs/hep-ph/0603175}{{\tt
  arXiv:hep-ph/0603175}}\relax
\mciteBstWouldAddEndPuncttrue
\mciteSetBstMidEndSepPunct{\mcitedefaultmidpunct}
{\mcitedefaultendpunct}{\mcitedefaultseppunct}\relax
\EndOfBibitem
\bibitem{LHCb-PROC-2010-056}
I.~Belyaev {\em et~al.}, \ifthenelse{\boolean{articletitles}}{{\it {Handling of
  the generation of primary events in \gauss, the \lhcb simulation framework}},
  }{}\href{http://dx.doi.org/10.1109/NSSMIC.2010.5873949}{Nuclear Science
  Symposium Conference Record (NSS/MIC) {\bf IEEE} (2010) 1155}\relax
\mciteBstWouldAddEndPuncttrue
\mciteSetBstMidEndSepPunct{\mcitedefaultmidpunct}
{\mcitedefaultendpunct}{\mcitedefaultseppunct}\relax
\EndOfBibitem
\bibitem{Lange:2001uf}
D.~J. Lange, \ifthenelse{\boolean{articletitles}}{{\it {The \evtgen~particle
  decay simulation package}},
  }{}\href{http://dx.doi.org/10.1016/S0168-9002(01)00089-4}{Nucl.\ Instrum.\
  Meth.\  {\bf A462} (2001) 152}\relax
\mciteBstWouldAddEndPuncttrue
\mciteSetBstMidEndSepPunct{\mcitedefaultmidpunct}
{\mcitedefaultendpunct}{\mcitedefaultseppunct}\relax
\EndOfBibitem
\bibitem{Golonka:2005pn}
P.~Golonka and Z.~Was, \ifthenelse{\boolean{articletitles}}{{\it {\photos~Monte
  Carlo: a precision tool for QED corrections in $\Z$ and $\W$ decays}},
  }{}\href{http://dx.doi.org/10.1140/epjc/s2005-02396-4}{Eur.\ Phys.\ J.\  {\bf
  C45} (2006) 97}, \href{http://arxiv.org/abs/hep-ph/0506026}{{\tt
  arXiv:hep-ph/0506026}}\relax
\mciteBstWouldAddEndPuncttrue
\mciteSetBstMidEndSepPunct{\mcitedefaultmidpunct}
{\mcitedefaultendpunct}{\mcitedefaultseppunct}\relax
\EndOfBibitem
\bibitem{Allison:2006ve}
GEANT4 collaboration, J.~Allison {\em et~al.},
  \ifthenelse{\boolean{articletitles}}{{\it {\geant developments and
  applications}}, }{}\href{http://dx.doi.org/10.1109/TNS.2006.869826}{IEEE
  Trans.\ Nucl.\ Sci.\  {\bf 53} (2006) 270}\relax
\mciteBstWouldAddEndPuncttrue
\mciteSetBstMidEndSepPunct{\mcitedefaultmidpunct}
{\mcitedefaultendpunct}{\mcitedefaultseppunct}\relax
\EndOfBibitem
\bibitem{Agostinelli:2002hh}
GEANT4 collaboration, S.~Agostinelli {\em et~al.},
  \ifthenelse{\boolean{articletitles}}{{\it {\geant: a simulation toolkit}},
  }{}\href{http://dx.doi.org/10.1016/S0168-9002(03)01368-8}{Nucl.\ Instrum.\
  Meth.\  {\bf A506} (2003) 250}\relax
\mciteBstWouldAddEndPuncttrue
\mciteSetBstMidEndSepPunct{\mcitedefaultmidpunct}
{\mcitedefaultendpunct}{\mcitedefaultseppunct}\relax
\EndOfBibitem
\bibitem{LHCb-PROC-2011-006}
M.~Clemencic {\em et~al.}, \ifthenelse{\boolean{articletitles}}{{\it {The \lhcb
  simulation application, \gauss: design, evolution and experience}},
  }{}\href{http://dx.doi.org/10.1088/1742-6596/331/3/032023}{{J.\ of Phys.\
  Conf.\ Ser.\ } {\bf 331} (2011) 032023}\relax
\mciteBstWouldAddEndPuncttrue
\mciteSetBstMidEndSepPunct{\mcitedefaultmidpunct}
{\mcitedefaultendpunct}{\mcitedefaultseppunct}\relax
\EndOfBibitem
\bibitem{LHCb-2008-002}
M.~Needham, \ifthenelse{\boolean{articletitles}}{{\it {Clone track
  identification using the Kullback-Leibler distance}}, }{}
  \href{http://cdsweb.cern.ch/search?p=CERN-LHCb-2008-002&f=reportnumber&action_search=Search&c=LHCb+Reports&c=LHCb+Conference+Proceedings&c=LHCb+Conference+Contributions&c=LHCb+Notes&c=LHCb+Theses&c=LHCb+Papers}
  {CERN-LHCb-2008-002}\relax
\mciteBstWouldAddEndPuncttrue
\mciteSetBstMidEndSepPunct{\mcitedefaultmidpunct}
{\mcitedefaultendpunct}{\mcitedefaultseppunct}\relax
\EndOfBibitem
\bibitem{Kullback1}
S.~Kullback and R.~A. Leibler, \ifthenelse{\boolean{articletitles}}{{\it {On
  information and sufficiency}}, }{}Annals of Mathematical Statistics {\bf 22}
  (1951) 79\relax
\mciteBstWouldAddEndPuncttrue
\mciteSetBstMidEndSepPunct{\mcitedefaultmidpunct}
{\mcitedefaultendpunct}{\mcitedefaultseppunct}\relax
\EndOfBibitem
\bibitem{Kullback3}
S.~Kullback, \ifthenelse{\boolean{articletitles}}{{\it {Letter to editor: the
  Kullback-Leibler distance}}, }{}The American Statistician {\bf 41} (1987)
  340\relax
\mciteBstWouldAddEndPuncttrue
\mciteSetBstMidEndSepPunct{\mcitedefaultmidpunct}
{\mcitedefaultendpunct}{\mcitedefaultseppunct}\relax
\EndOfBibitem
\bibitem{Muon:performance}
A.~A. Alves {\em et~al.}, \ifthenelse{\boolean{articletitles}}{{\it
  {Performance of the LHCb muon system}},
  }{}\href{http://dx.doi.org/10.1088/1748-0221/8/02/P02022}{JINST {\bf 8}
  (2013) P02022}, \href{http://arxiv.org/abs/1211.1346}{{\tt
  arXiv:1211.1346}}\relax
\mciteBstWouldAddEndPuncttrue
\mciteSetBstMidEndSepPunct{\mcitedefaultmidpunct}
{\mcitedefaultendpunct}{\mcitedefaultseppunct}\relax
\EndOfBibitem
\bibitem{arXiv:1211-6759}
M.~Adinolfi {\em et~al.}, \ifthenelse{\boolean{articletitles}}{{\it
  {Performance of the \lhcb RICH detector at the LHC}},
  }{}\href{http://arxiv.org/abs/1211.6759}{{\tt arXiv:1211.6759}}, {submitted
  to Eur. Phys. J.}\relax
\mciteBstWouldAddEndPunctfalse
\mciteSetBstMidEndSepPunct{\mcitedefaultmidpunct}
{}{\mcitedefaultseppunct}\relax
\EndOfBibitem
\bibitem{LHCb-PAPER-2012-022}
LHCb collaboration, R.~Aaij {\em et~al.},
  \ifthenelse{\boolean{articletitles}}{{\it {Evidence for the decay
  $\Bd\to\jpsi\Pomega$ and measurement of the relative branching fractions of
  \Bs~meson decays to $\jpsi\Peta$ and $\jpsi\Peta^\prime$}},
  }{}\href{http://dx.doi.org/10.1016/j.nuclphysb.2012.10.021}{Nucl.\ Phys.\
  {\bf B867} (2013) 547}, \href{http://arxiv.org/abs/1210.2631}{{\tt
  arXiv:1210.2631}}\relax
\mciteBstWouldAddEndPuncttrue
\mciteSetBstMidEndSepPunct{\mcitedefaultmidpunct}
{\mcitedefaultendpunct}{\mcitedefaultseppunct}\relax
\EndOfBibitem
\bibitem{LHCb-PAPER-2012-010}
LHCb collaboration, R.~Aaij {\em et~al.},
  \ifthenelse{\boolean{articletitles}}{{\it {Measurement of relative branching
  fractions of $\B$~decays to $\psitwos$ and $\jpsi$~mesons}},
  }{}\href{http://dx.doi.org/10.1140/epjc/s10052-012-2118-7}{Eur.\ Phys.\ J.\
  {\bf C72} (2012) 2118}, \href{http://arxiv.org/abs/1205.0918}{{\tt
  arXiv:1205.0918}}\relax
\mciteBstWouldAddEndPuncttrue
\mciteSetBstMidEndSepPunct{\mcitedefaultmidpunct}
{\mcitedefaultendpunct}{\mcitedefaultseppunct}\relax
\EndOfBibitem
\bibitem{LHCb-PAPER-2012-053}
LHCb collaboration, R.~Aaij {\em et~al.},
  \ifthenelse{\boolean{articletitles}}{{\it {Observation of the $\Bs\to
  \psitwos\Peta$ and $\B^0_{(\mathrm{s})} \to \psitwos\pipi$~decays}},
  }{}\href{http://dx.doi.org/10.1016/j.nuclphysb.2013.03.004}{Nucl.\ Phys.\
  {\bf B871} (2013) 403}, \href{http://arxiv.org/abs/1302.6354}{{\tt
  arXiv:1302.6354}}\relax
\mciteBstWouldAddEndPuncttrue
\mciteSetBstMidEndSepPunct{\mcitedefaultmidpunct}
{\mcitedefaultendpunct}{\mcitedefaultseppunct}\relax
\EndOfBibitem
\bibitem{LHCb-PAPER-2012-003}
LHCb collaboration, R.~Aaij {\em et~al.},
  \ifthenelse{\boolean{articletitles}}{{\it {Observation of double charm
  production involving open charm in $\mathrm{pp}$~collisions at $\sqrt{s}=
  7\tev$}}, }{}\href{http://dx.doi.org/10.1007/JHEP06(2012)141}{JHEP {\bf 06}
  (2012) 141}, \href{http://arxiv.org/abs/1205.0975}{{\tt
  arXiv:1205.0975}}\relax
\mciteBstWouldAddEndPuncttrue
\mciteSetBstMidEndSepPunct{\mcitedefaultmidpunct}
{\mcitedefaultendpunct}{\mcitedefaultseppunct}\relax
\EndOfBibitem
\bibitem{Hulsbergen:2005pu}
W.~D. Hulsbergen, \ifthenelse{\boolean{articletitles}}{{\it {Decay chain
  fitting with a Kalman filter}},
  }{}\href{http://dx.doi.org/10.1016/j.nima.2005.06.078}{Nucl.\ Instrum.\
  Meth.\  {\bf A552} (2005) 566},
  \href{http://arxiv.org/abs/physics/0503191}{{\tt
  arXiv:physics/0503191}}\relax
\mciteBstWouldAddEndPuncttrue
\mciteSetBstMidEndSepPunct{\mcitedefaultmidpunct}
{\mcitedefaultendpunct}{\mcitedefaultseppunct}\relax
\EndOfBibitem
\bibitem{LHCb-PAPER-2013-011}
LHCb collaboration, R.~Aaij {\em et~al.},
  \ifthenelse{\boolean{articletitles}}{{\it {Precision measurement of \D~meson
  mass differences}}, }{}\href{http://arxiv.org/abs/1304.6865}{{\tt
  arXiv:1304.6865}}\relax
\mciteBstWouldAddEndPuncttrue
\mciteSetBstMidEndSepPunct{\mcitedefaultmidpunct}
{\mcitedefaultendpunct}{\mcitedefaultseppunct}\relax
\EndOfBibitem
\bibitem{LHCb-PAPER-2012-028}
LHCb collaboration, R.~Aaij {\em et~al.},
  \ifthenelse{\boolean{articletitles}}{{\it {Measurements of \Bc~production and
  mass with the $\Bc\to\jpsi\pip$ decay}},
  }{}\href{http://dx.doi.org/10.1103/PhysRevLett.109.232001}{Phys.\ Rev.\
  Lett.\  {\bf 109} (2012) 232001}, \href{http://arxiv.org/abs/1209.5634}{{\tt
  arXiv:1209.5634}}\relax
\mciteBstWouldAddEndPuncttrue
\mciteSetBstMidEndSepPunct{\mcitedefaultmidpunct}
{\mcitedefaultendpunct}{\mcitedefaultseppunct}\relax
\EndOfBibitem
\bibitem{Skwarnicki:1986xj}
T.~Skwarnicki, {\em {A study of the radiative cascade transitions between the
  $\Upsilon^{\prime}$~and $\Upsilon$~resonances}}, PhD thesis, Institute of
  Nuclear Physics, Krakow, 1986,
  {\href{http://inspirehep.net/record/230779/files/230779.pdf}{DESY-F31-86-02}}\relax
\mciteBstWouldAddEndPuncttrue
\mciteSetBstMidEndSepPunct{\mcitedefaultmidpunct}
{\mcitedefaultendpunct}{\mcitedefaultseppunct}\relax
\EndOfBibitem
\bibitem{LHCb-PAPER-2011-013}
LHCb collaboration, R.~Aaij {\em et~al.},
  \ifthenelse{\boolean{articletitles}}{{\it {Observation of \jpsi~pair
  production in $\mathrm{pp}$~collisions at $\sqrt{s}=7\tev$}},
  }{}\href{http://dx.doi.org/10.1016/j.physletb.2011.12.015}{Phys.\ Lett.\
  {\bf B707} (2012) 52}, \href{http://arxiv.org/abs/1109.0963}{{\tt
  arXiv:1109.0963}}\relax
\mciteBstWouldAddEndPuncttrue
\mciteSetBstMidEndSepPunct{\mcitedefaultmidpunct}
{\mcitedefaultendpunct}{\mcitedefaultseppunct}\relax
\EndOfBibitem
\bibitem{Lees:2011gw}
\babar~collaboration, J.-P. Lees {\em et~al.},
  \ifthenelse{\boolean{articletitles}}{{\it {Branching fraction measurements of
  the color-suppressed decays $\bar{\mathrm{B}}^0 \to
  \mathrm{D}^{\left(*\right)0} \Ppi^0$, $\mathrm{D}^{\left(*\right)0} \Peta$,
  $\mathrm{D}^{\left(*\right)0} \Pomega$, and $\mathrm{D}^{\left(*\right)0}
  \Peta^{\prime}$ and measurement of the polarization in the decay
  $\bar{\mathrm{B}}^0 \to \mathrm{D}^{*0} \Pomega$}},
  }{}\href{http://dx.doi.org/10.1103/PhysRevD.84.112007}{Phys.\ Rev.\  {\bf
  D84} (2011) 112007}, \href{http://arxiv.org/abs/1107.5751}{{\tt
  arXiv:1107.5751}}\relax
\mciteBstWouldAddEndPuncttrue
\mciteSetBstMidEndSepPunct{\mcitedefaultmidpunct}
{\mcitedefaultendpunct}{\mcitedefaultseppunct}\relax
\EndOfBibitem
\bibitem{Lyons:2008}
L.~Lyons, \ifthenelse{\boolean{articletitles}}{{\it {Open statistical issues in
  Particle Physics}}, }{}Annals of Applied Statistics {\bf 2} (2008) 887\relax
\mciteBstWouldAddEndPuncttrue
\mciteSetBstMidEndSepPunct{\mcitedefaultmidpunct}
{\mcitedefaultendpunct}{\mcitedefaultseppunct}\relax
\EndOfBibitem
\bibitem{Gross:2010}
E.~Gross and O.~Vitells, \ifthenelse{\boolean{articletitles}}{{\it Trial
  factors for the look elsewhere effect in high energy physics},
  }{}\href{http://dx.doi.org/10.1140/epjc/s10052-010-1470-8}{Eur.\ Phys.\ J.\
  {\bf C70} (2010) 525}\relax
\mciteBstWouldAddEndPuncttrue
\mciteSetBstMidEndSepPunct{\mcitedefaultmidpunct}
{\mcitedefaultendpunct}{\mcitedefaultseppunct}\relax
\EndOfBibitem
\bibitem{LHCb-PAPER-2012-048}
LHCb collaboration, R.~Aaij {\em et~al.},
  \ifthenelse{\boolean{articletitles}}{{\it {Measurements of the
  $\Lambda_{\bquark}^0$, $\Xi_{\bquark}^-$ and $\Omega_{\bquark}^-$ baryon
  masses}}, }{}\href{http://arxiv.org/abs/1302.1072}{{\tt arXiv:1302.1072}}, to
  appear in Phys. Rev. Lett.\relax
\mciteBstWouldAddEndPunctfalse
\mciteSetBstMidEndSepPunct{\mcitedefaultmidpunct}
{}{\mcitedefaultseppunct}\relax
\EndOfBibitem
\bibitem{LHCb-PAPER-2011-035}
LHCb collaboration, R.~Aaij {\em et~al.},
  \ifthenelse{\boolean{articletitles}}{{\it {Measurement of \bquark-hadron
  masses}}, }{}\href{http://dx.doi.org/10.1016/j.physletb.2012.01.058}{Phys.\
  Lett.\  {\bf B708} (2012) 241}, \href{http://arxiv.org/abs/1112.4896}{{\tt
  arXiv:1112.4896}}\relax
\mciteBstWouldAddEndPuncttrue
\mciteSetBstMidEndSepPunct{\mcitedefaultmidpunct}
{\mcitedefaultendpunct}{\mcitedefaultseppunct}\relax
\EndOfBibitem
\bibitem{Pivk:2004ty}
M.~Pivk and F.~R. Le~Diberder, \ifthenelse{\boolean{articletitles}}{{\it
  {\sPlot{}: a Statistical tool to unfold data distributions}},
  }{}\href{http://dx.doi.org/10.1016/j.nima.2005.08.106}{Nucl.\ Instrum.\
  Meth.\  {\bf A555} (2005) 356},
  \href{http://arxiv.org/abs/physics/0402083}{{\tt
  arXiv:physics/0402083}}\relax
\mciteBstWouldAddEndPuncttrue
\mciteSetBstMidEndSepPunct{\mcitedefaultmidpunct}
{\mcitedefaultendpunct}{\mcitedefaultseppunct}\relax
\EndOfBibitem
\bibitem{LHCb-PAPER-2010-001}
LHCb collaboration, R.~Aaij {\em et~al.},
  \ifthenelse{\boolean{articletitles}}{{\it {Prompt $\KS$~production in
  $\mathrm{pp}$~collisions at \mbox{$\sqrt{s}=0.9\tev$}}},
  }{}\href{http://dx.doi.org/10.1016/j.physletb.2010.08.055}{Phys.\ Lett.\
  {\bf B693} (2010) 69}, \href{http://arxiv.org/abs/1008.3105}{{\tt
  arXiv:1008.3105}}\relax
\mciteBstWouldAddEndPuncttrue
\mciteSetBstMidEndSepPunct{\mcitedefaultmidpunct}
{\mcitedefaultendpunct}{\mcitedefaultseppunct}\relax
\EndOfBibitem
\bibitem{LHCb-PUB-2011-025}
A.~Jaeger {\em et~al.}, \ifthenelse{\boolean{articletitles}}{{\it {Measurement
  of the track finding efficiency}}, }{}
  \href{http://cdsweb.cern.ch/search?p=LHCb-PUB-2011-025&f=reportnumber&action_search=Search&c=LHCb+Reports&c=LHCb+Conference+Proceedings&c=LHCb+Conference+Contributions&c=LHCb+Notes&c=LHCb+Theses&c=LHCb+Papers}
  {LHCb-PUB-2011-025}\relax
\mciteBstWouldAddEndPuncttrue
\mciteSetBstMidEndSepPunct{\mcitedefaultmidpunct}
{\mcitedefaultendpunct}{\mcitedefaultseppunct}\relax
\EndOfBibitem
\bibitem{Alexander:2008aa}
CLEO collaboration, J.~Alexander {\em et~al.},
  \ifthenelse{\boolean{articletitles}}{{\it {Absolute measurement of hadronic
  branching fractions of the \Ds~meson}},
  }{}\href{http://dx.doi.org/10.1103/PhysRevLett.100.161804}{Phys.\ Rev.\
  Lett.\  {\bf 100} (2008) 161804}, \href{http://arxiv.org/abs/0801.0680}{{\tt
  arXiv:0801.0680}}\relax
\mciteBstWouldAddEndPuncttrue
\mciteSetBstMidEndSepPunct{\mcitedefaultmidpunct}
{\mcitedefaultendpunct}{\mcitedefaultseppunct}\relax
\EndOfBibitem
\bibitem{Ahmed:2000ad}
CLEO collaboration, S.~Ahmed {\em et~al.},
  \ifthenelse{\boolean{articletitles}}{{\it {Measurement of
  $\Bd\to\D_{\squark}^{(\ast)+}\D^{\ast(\ast)}$ branching fractions}},
  }{}\href{http://dx.doi.org/10.1103/PhysRevD.62.112003}{Phys.\ Rev.\  {\bf
  D62} (2000) 112003}, \href{http://arxiv.org/abs/hep-ex/0008015}{{\tt
  arXiv:hep-ex/0008015}}\relax
\mciteBstWouldAddEndPuncttrue
\mciteSetBstMidEndSepPunct{\mcitedefaultmidpunct}
{\mcitedefaultendpunct}{\mcitedefaultseppunct}\relax
\EndOfBibitem
\bibitem{Aubert:2003jj}
\babar collaboration, B.~Aubert {\em et~al.},
  \ifthenelse{\boolean{articletitles}}{{\it {Measurement of
  $\Bd\to\D_{\squark}^{(\ast)+}\D^{(\ast)-}$ branching fractions and
  $\Bd\to\D_{\squark}^{\ast+}\D^{\ast-}$ polarization with a partial
  reconstruction technique}},
  }{}\href{http://dx.doi.org/10.1103/PhysRevD.67.092003}{Phys.\ Rev.\  {\bf
  D67} (2003) 092003}, \href{http://arxiv.org/abs/hep-ex/0302015}{{\tt
  arXiv:hep-ex/0302015}}\relax
\mciteBstWouldAddEndPuncttrue
\mciteSetBstMidEndSepPunct{\mcitedefaultmidpunct}
{\mcitedefaultendpunct}{\mcitedefaultseppunct}\relax
\EndOfBibitem
\bibitem{Richman}
J.~D. Richman, in {\em {Probing the Standard Model of particle interactions}},
  {R.~Gupta, A.~Morel, E.~de~Rafael and F.~David}, ed., p.~640, {Elsevier},
  {Amsterdam}, 1999\relax
\mciteBstWouldAddEndPuncttrue
\mciteSetBstMidEndSepPunct{\mcitedefaultmidpunct}
{\mcitedefaultendpunct}{\mcitedefaultseppunct}\relax
\EndOfBibitem
\bibitem{Ebert:2003cn}
D.~Ebert, R.~Faustov, and V.~Galkin, \ifthenelse{\boolean{articletitles}}{{\it
  {Weak decays of the \Bc~meson to charmonium and \D~mesons in the relativistic
  quark model}}, }{}\href{http://dx.doi.org/10.1103/PhysRevD.68.094020}{Phys.\
  Rev.\  {\bf D68} (2003) 094020},
  \href{http://arxiv.org/abs/hep-ph/0306306}{{\tt arXiv:hep-ph/0306306}}\relax
\mciteBstWouldAddEndPuncttrue
\mciteSetBstMidEndSepPunct{\mcitedefaultmidpunct}
{\mcitedefaultendpunct}{\mcitedefaultseppunct}\relax
\EndOfBibitem
\bibitem{Likhoded:2009ib}
A.~Likhoded and A.~Luchinsky, \ifthenelse{\boolean{articletitles}}{{\it {Light
  hadron production in $\Bc\to\jpsi\mathrm{X}$~decays}},
  }{}\href{http://dx.doi.org/10.1103/PhysRevD.81.014015}{Phys.\ Rev.\  {\bf
  D81} (2010) 014015}, \href{http://arxiv.org/abs/0910.3089}{{\tt
  arXiv:0910.3089}}\relax
\mciteBstWouldAddEndPuncttrue
\mciteSetBstMidEndSepPunct{\mcitedefaultmidpunct}
{\mcitedefaultendpunct}{\mcitedefaultseppunct}\relax
\EndOfBibitem
\end{mcitethebibliography}

\end{document}